\documentclass[twocolumn,superscriptaddress,floatfix]{revtex4}

\usepackage{multirow,amssymb,amsbsy,amsmath}
\usepackage{graphicx}
\usepackage{verbatim}
\makeatletter
\def\bra#1{\mathinner{\langle{#1}|}}
\def\ket#1{\mathinner{|{#1}\rangle}}

\usepackage{color}

\begin{document}

\title{Photonic realization of nonlocal memory effects and non-Markovian quantum probes}

\author{Bi-Heng Liu}
\affiliation{Key Laboratory of Quantum Information, University of Science and 
Technology of China, CAS, Hefei, 230026, China}

\author{Dong-Yang Cao}
\affiliation{Key Laboratory of Quantum Information, University of Science and 
Technology of China, CAS, Hefei, 230026, China}

\author{Yun-Feng Huang }
\affiliation{Key Laboratory of Quantum Information, University of Science and 
Technology of China, CAS, Hefei, 230026, China}

\author{Chuan-Feng Li}
\email{cfli@ustc.edu.cn}
\affiliation{Key Laboratory of Quantum Information, University of Science and 
Technology of China, CAS, Hefei, 230026, China}

\author{Guang-Can Guo}
\affiliation{Key Laboratory of Quantum Information, University of Science and 
Technology of China, CAS, Hefei, 230026, China}

\author{Elsi-Mari Laine}
\affiliation{Turku Centre for Quantum Physics, Department of Physics and 
Astronomy, University of Turku, FI-20014 Turun yliopisto, Finland}

\author{Heinz-Peter Breuer}
\affiliation{Physikalisches Institut, Universit\"at Freiburg,
Hermann-Herder-Strasse 3, D-79104 Freiburg, Germany}

\author{Jyrki Piilo}
\email{jyrki.piilo@utu.fi}
\affiliation{Turku Centre for Quantum Physics, Department of Physics and 
Astronomy, University of Turku, FI-20014 Turun yliopisto, Finland}

\date{\today}

\begin{abstract}
The study of open quantum systems is important for fundamental issues of quantum physics as well as for technological applications such as quantum information processing. Recent developments in this field have increased our basic understanding on how non-Markovian effects influence the dynamics of an open quantum system, paving the way to exploit memory effects for various quantum control tasks. Most often, the environment of an open system is thought to act as a sink for the system information. However, here we demonstrate
experimentally that a photonic open system can exploit the 
information initially held by its environment. Correlations in the 
environmental degrees of freedom induce nonlocal memory effects where the 
bipartite open system displays, counterintuitively, local Markovian and global non-Markovian 
character. Our results also provide novel methods to protect and distribute
entanglement, and to experimentally quantify correlations in photonic 
environments.

\end{abstract}

\maketitle

\section*{Introduction}

Realistic quantum systems interact and exchange information 
with their surroundings. The engineering of the decoherence and the flow of 
information between an open quantum system and its environment has recently 
allowed, e.g., to drive quantum computation by dissipation~\cite{Verstraete}, to 
control entanglement and quantum phases in many-body 
systems~\cite{Diehl,Cho,Krauter}, to create an open system quantum 
simulator~\cite{Barreiro}, and to control the transition from Markovian
dynamics to a regime with quantum memory effects~\cite{Liu}. 
Moreover, there has been recently rapid progress in understanding various fundamental issues on quantum non-Markovianity~\cite{Piilo08,Vacchini08,Wolf,Lidar2009,BLP,RHP,Kossakowski2010}.

In general, the power of quantum information processing is based on the presence of 
quantum correlations in multipartite systems. However, as any 
realistic quantum system, those systems are vulnerable to the 
detrimental effects of the environment such as the loss of coherence and 
information. Depending on its physical composition, an open multipartite system 
can interact either with a single common environment or each of the sub-parties 
can interact locally with their own reservoirs. The latter case, which is in the focus 
of our paper, is a common scenario, e.g., in quantum information processing and 
entanglement dynamics~\cite{Bellomo}, and in energy transport in biological 
organisms~\cite{Rebentrost}.

The central question to be studied here is, can a composite open quantum system
exploit during its time evolution the information carried initially by the 
environmental degrees of freedom, i.e., the information to which it does not have 
access initially? On the basis of a recent theoretical prediction~\cite{Laine11}, we
demonstrate experimentally that initial correlations between local parts of the
environment lead to nonlocal memory effects 
and allow to protect and maintain genuine quantum properties of an
open system. We will show further that the non-Markovian dynamics
of photonic open systems provides a diagnostic tool for characteristic
features of their environments.

\section*{Results}

\subsection*{The physical system and experimental setup}

Our experimental open system consists of the polarization degrees of freedom of a 
pair of entangled photons. The photon pair is created by spontaneous 
parametric down-conversion after which the photons travel along different 
arms $i=1,2$ and move through different quartz plates of variable thickness (see
Fig.~1). The quartz plates act as birefringent media which lead to a 
local coupling between the polarization degrees of freedom of the photon in arm 
$i$ and its frequency (mode) degrees of freedom forming its local environment.
This local interaction can be described by a local unitary operator
which is defined by 
$U_i(t_i)|\lambda\rangle\otimes|\omega_i\rangle=e^{in_{\lambda}\omega_i t_i}
|\lambda\rangle\otimes|\omega_i\rangle$, where 
$|\lambda \rangle \otimes |\omega_i\rangle$ denotes the state of the 
photon in arm $i$ with polarization $\lambda = H,V$ (horizontal or vertical) and 
frequency $\omega_i$. The refraction index for photons with polarization 
$\lambda$ is denoted by $n_{\lambda}$, and $t_i$ represents the interaction time.

\begin{figure}[tb]
\centering
\includegraphics[width=0.4\textwidth]{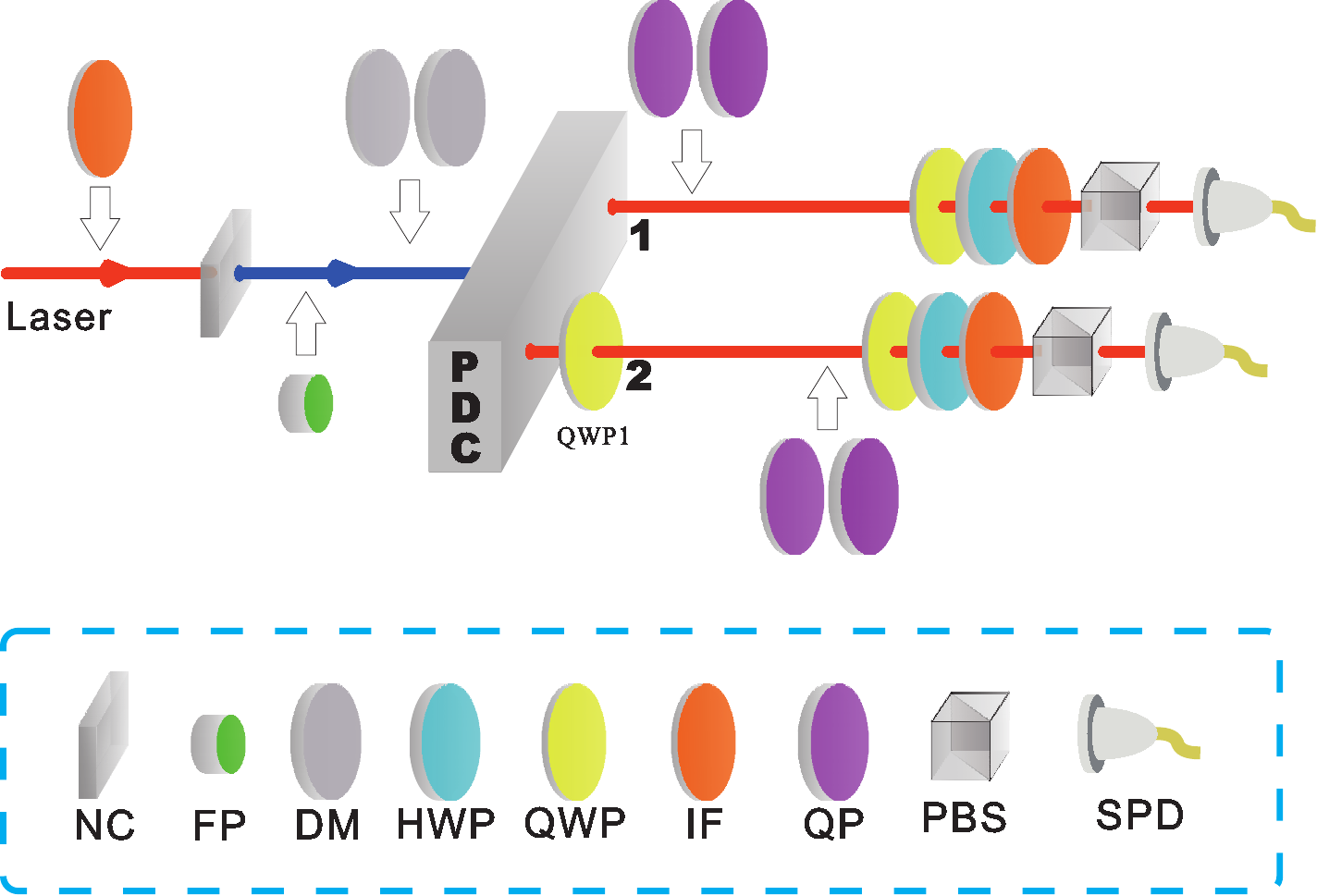}
\caption{\label{Fig:1} 
Experimental setup. Maximally entangled photon pairs are generated by 
parametric down conversion (PDC) with an ultraviolet (UV) pump pulse whose 
spectral width can be controlled. Then a series of quartz plates are added in arm 
$1$ and arm $2$ to realize the local dephasing channels. The final two-photon 
polarization state is analyzed by state tomography. Key to the components: 
NC -- nonlinear crystal, FP -- Fabry-P\'erot cavity, DM -- dispersion medium, HWP -- half wave plate, QWP -- quarter wave plate, IF -- 
interference filter, QP -- quartz plate, PBS -- polarizing beamsplitter, and  SPD -- 
single photon detector.}
\end{figure}

Any pure initial state for the polarization degrees of freedom of the photon pair, 
which forms our bipartite open system, can be written as 
$\ket{\psi_{12}}=a \ket{HH}+b\ket{HV}+c\ket{VH}+d\ket{VV}$. The total 
system initial states are assumed to be product states
\begin{equation} \label{total}
 \ket{\Psi(0)} = \ket{\psi_{12}} \otimes \int d\omega_1 d\omega_2 \,
 g(\omega_1,\omega_2) \ket{\omega_1,\omega_2}, \nonumber
\end{equation}
where $g(\omega_1,\omega_2)$ is the probability amplitude for the photon in
arm $1$ to have frequency $\omega_1$ and for the photon in arm $2$ to have 
frequency $\omega_2$, with the corresponding joint probability distribution  
$P(\omega_1,\omega_2)=|g(\omega_1,\omega_2)|^2$. 

\subsection*{Open system dynamics}

The time evolution of the open system can be represented by means of the
dynamical map $\Phi^t_{12}$ \cite{Breuer2007} which maps the initial polarization 
state $\rho_{12}(0) = \ket{\psi_{12}}\bra{\psi_{12}}$ to the polarization state state 
$\rho_{12}(t) = \Phi^t_{12} \big(\rho_{12}(0)\big)$ at time $t$.
Calculating the total system time evolution and taking the trace over the 
environment yields~\cite{Laine11}
\begin{equation} \label{rho12}
\rho_{12}(t)  =\left(\begin{array}{cccc}
 |a|^2 & a b^* \kappa_2(t) & ac^* \kappa_1(t)& ad^* \kappa_{12}(t) \\
 b a^* \kappa_2^*(t) & |b|^2 & b c^*\Lambda_{12}(t) & b d^* \kappa_1(t)\\
 c a^* \kappa_1^*(t) & c b^* \Lambda_{12}^*(t) & |c|^2 & c d^* \kappa_2(t)\\
 d a^* \kappa_{12}^*(t) & d b^* \kappa_1^*(t) & d c^* \kappa_2^*(t) & |d|^2
\end{array}\right).
\end{equation}
The dynamics represents a pure dephasing process whose decoherence
functions can be expressed completely in terms of the Fourier transform of
the joint probability distribution 
\begin{equation} \label{Gt1t2}
 G(t_1,t_2)=\int d\omega_1 d\omega_2 P(\omega_1,\omega_2) 
 e^{-i\Delta n(\omega_1t_1+\omega_2t_2)}
\end{equation}
as $\kappa_1(t)=G(t_1,0)$, $\kappa_2(t)=G(0,t_2)$, 
$\kappa_{12}(t)=G(t_1,t_2)$ and $\Lambda_{12}(t)=G(t_1,-t_2)$, where
$\Delta n = n_V - n_H$ denotes the birefringence. The function $\kappa_i(t)$ 
describes the decoherence of the polarization state of photon $i$ which is
induced by the local coupling to its frequency environment caused by the
quartz plate in arm $i$. This can be seen by taking the trace over the polarization
state of photon 2 to obtain the polarization state of photon 1: 
\begin{equation}
 \rho_{1}(t)= \left(\begin{array}{cc}
 |a|^2+|b|^2 & (a c^*+b d^*) \kappa_1(t)  \\
 (c a^*+d b^*) \kappa_1^*(t) & |c|^2+|d|^2.
\end{array}\right),\nonumber
\end{equation}
or by tracing over photon 1 to find the state of photon 2:
\begin{equation}
 \rho_{2}(t)=\left(\begin{array}{cc}
 |a|^2+|c|^2 & (a b^*+c d^*) \kappa_2(t)  \\
 (b a^*+d c^*) \kappa_2^*(t) & |b|^2+|d|^2
\end{array}\right).\nonumber
\end{equation}

\subsection*{Nonlocal dynamical map}

The nonlocal character of the dynamical map $\Phi_{12}^t$ is controlled by
the decoherence functions $\kappa_{12}(t)$ and $\Lambda_{12}(t)$
in Eq.~(\ref{rho12}). As a matter of fact, the map $\Phi_{12}^t$ is a product 
of local dynamical maps, i.e., $\Phi_{12}^t=\Phi_1^t\otimes\Phi_2^t$ if and only if 
$\kappa_{12}(t)=\kappa_1(t)\kappa_2(t)$ and 
$\Lambda_{12}(t)= \kappa_1(t)\kappa^*_2(t)$. 
In particular, this implies that frequency correlations in the initial state do not 
influence the local dynamics of each photon and cannot be detected by
observing the local polarization states. However, the evolution of the composite 
two-photon polarization state is determined by the frequency correlations in the 
environment which lead to memory effects and global non-Markovian dynamics.
This also means that, counterintuitively, by adding degrees of freedom 
to an open system one can change its dynamics from the Markovian to 
the non-Markovian regime.

\subsection*{Trace distance dynamics}

To quantify non-Markovianity we use the trace distance based measure \cite{BLP} 
which is defined by
$\mathcal{N}(\Phi)
=\max_{\rho_{A,B}(0)}\int_{\sigma>0}dt\,\sigma(t,\rho_{ A,B}(0))$. Here, 
$\sigma(t,\rho_{A,B}(0))=\frac{d}{dt}D(\rho_A(t),\rho_B(t))$ and the trace 
distance $D$ between two states $\rho_A$ and $\rho_B$ is given by
$D(\rho_A, \rho_B) = \frac{1}{2} \textrm{tr}|\rho_A-\rho_B|$. 
The amount of non-Markovianity is therefore equal to the total increase of the 
trace distance during the time evolution and quantifies the total information flow 
from the environment to the system. This measure has been used recently in 
several theoretical and experimental contexts, see, 
e.g.~Refs.~\cite{Rebentrost,Liu,Paternostro,Tang}.

\begin{figure}[tb]
\centering
\includegraphics[width=0.5\textwidth]{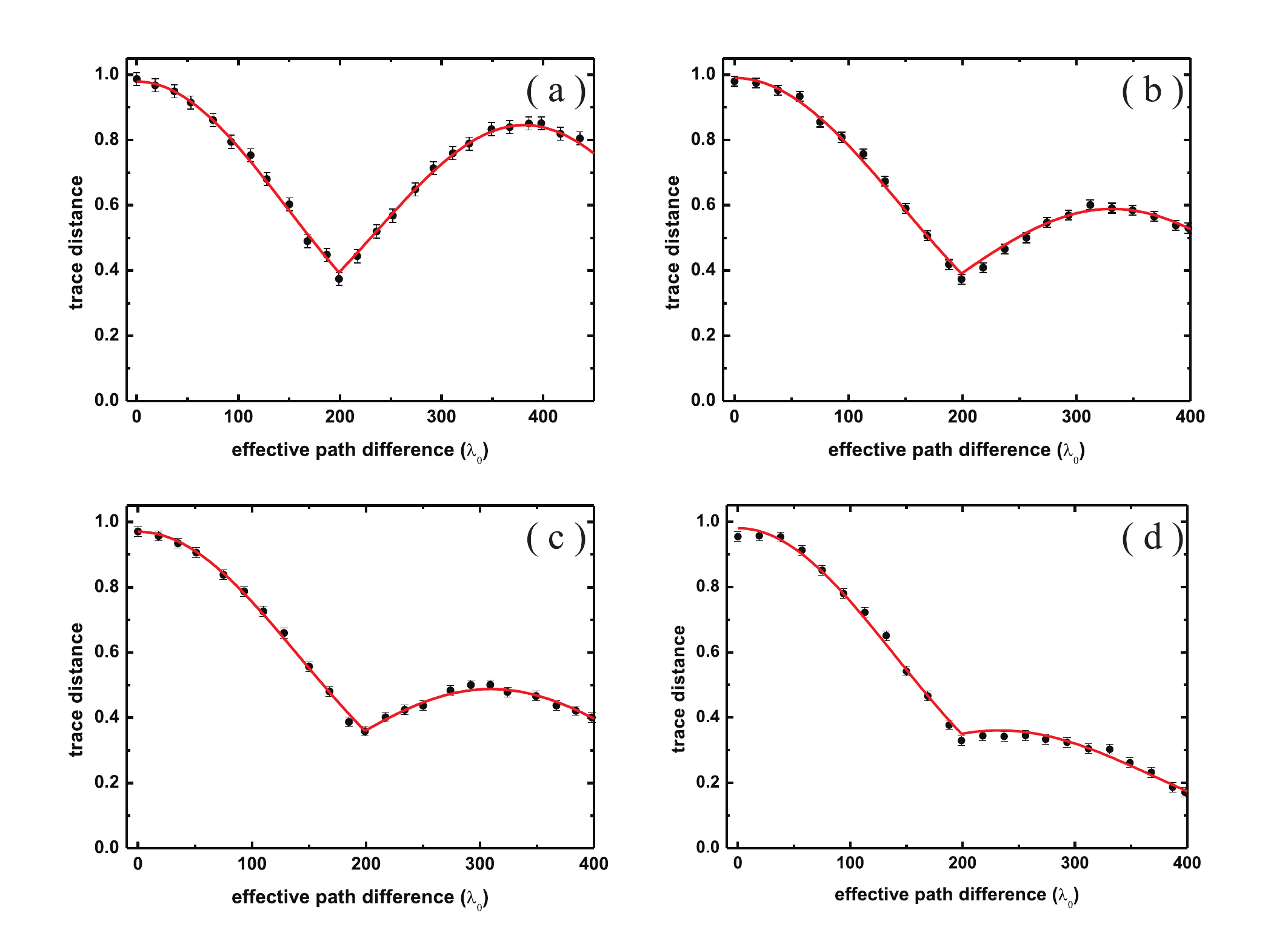}
\caption{\label{Fig:2} 
Trace distance dynamics for different values of the spectral width of the 
pump pulse. The widths and the amounts of non-Markovianity are:
(a) $\delta=0.18\,\rm{nm}$, $\mathcal{N} = 0.48$,
(b) $\delta=0.52\,\rm{nm}$, $\mathcal{N} = 0.23$,
(c) $\delta=0.73\,\rm{nm}$, $\mathcal{N} = 0.14$,
(d) $\delta=1.89\,\rm{nm}$, $\mathcal{N} = 0.02$.
The $x$-axis represents time measured by the total effective path difference 
between the horizontal and vertical photons caused by quartz plates in arms $1$ 
and $2$ ($\lambda_0=780\,\rm{nm}$). The $y$-axis is the trace distance between the 
time-evolved initial open system states $|\psi_{12}^{+}\rangle$ and 
$|\psi_{12}^{-}\rangle$. We first add quartz plates in arm 1. When the total effective 
path difference between horizontal and vertical photons equals $199\lambda_0$, 
we add quartz plates in arm 2. The solid line shows the fit using the theoretical 
result of Eq.~(\ref{D}). The error bars are due to the counting statistics. See the 
Supplementary Information for more details on the counting statistics and the fitting of the theoretical curves.}
\end{figure}

For anticorrelated frequencies $\omega_i$ the pair of the Bell-states
$|\psi_{12}^{\pm}\rangle = \left( |HH\rangle \pm |VV\rangle\right) / \sqrt{2}$
maximizes the increase of the trace distance and thus determines the 
non-Markovianity measure $\mathcal{N}$~\cite{Laine11}. These states
are created by using spontaneous parametric down conversion as is illustrated in 
Fig.~1. A femtosecond pulse (the duration is about $150\,\rm{fs}$ and the 
operation wavelength is at $780\,\rm{nm}$, with a repetition rate of about 
$76\,\rm{MHz}$) generated from a Ti:sapphire laser is frequency doubled to 
pump two $1$mm-thick beamlike cut beta barium borate (BBO) crystals~\cite{Niu} 
creating the two-photon entangled state. With the help of QWP1 we can easily 
tune the entangled state between $\ket{\psi_{12}^{+}}$ (QWP1 is set at $0^{\circ}$)
and $\ket{\psi_{12}^{-}}$ (QWP1 is set at  $90^{\circ}$).
The size of the anticorrelations between the photon frequencies is controlled
in the experiment by the spectrum of the UV pump pulse. Since energy is 
conserved in the down conversion process, decreasing the spectral width of the 
pump pulse decreases the uncertainty of the sum of the frequencies of the 
photons, and hence increases the anticorrelation between the frequencies. 
We use four different pulse widths as described in more detail in the Methods Section.
It is also worth noting that even though our experimental setup is based on downconversion, which is a commonly used tool to prepare specific two-photon states in quantum optical experiments, the physical phenomena that we describe and realize experimentally was theoretically predicted only very recently~\cite{Laine11}.

\begin{figure}[tb]
\centering
\includegraphics[width=0.5\textwidth]{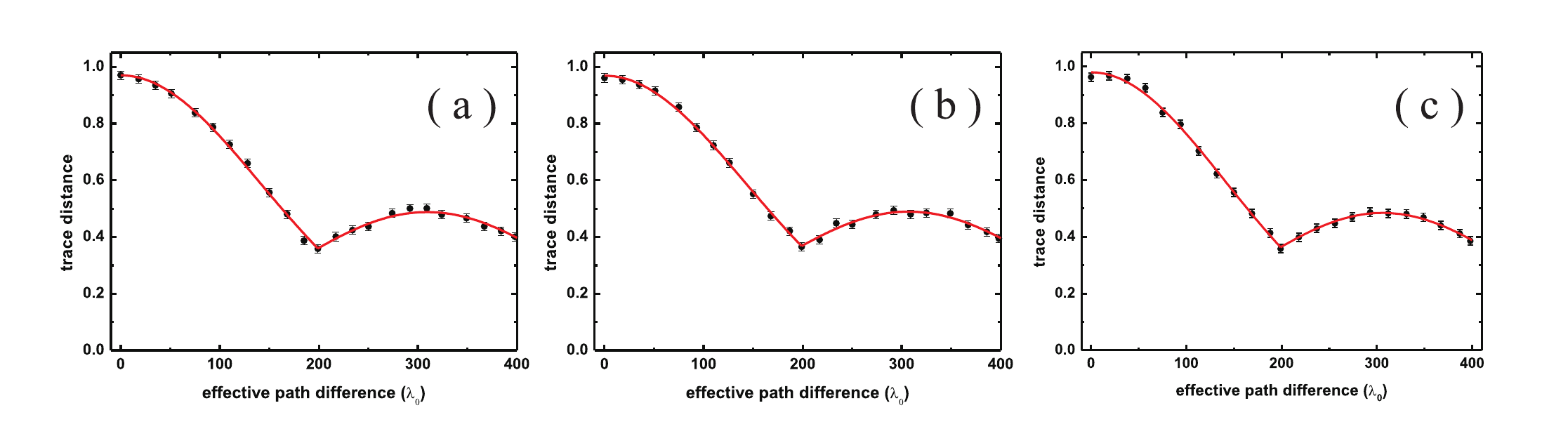}
\caption{\label{Fig:3} 
Trace distance dynamics of the open system with different dispersions of the pump 
pulse. The quartz plate on the path of the pump pulse has thickness
(a) $0$
(b) $13.19\,\rm{mm}$
(c) $39.57\,\rm{mm}$.
In all cases $K=-0.55$ and the measured values of non-Markovianity are 
$\mathcal{N} = 0.14, 0.13, 0.13$.
The trace distance dynamics remains the same irrespective of the character of 
the correlations. See the Methods Section for further details.}
\end{figure}

Assuming a Gaussian joint frequency distribution $P(\omega_1,\omega_2)$ 
with identical single frequency variance $C$ and correlation coefficient $K$,
the time evolution of the trace distance corresponding to the optimal
Bell-state pair is found to be~\cite{Laine11}
\begin{equation} \label{D}
 D(t)= \exp\left[
 -\frac{1}{2}\Delta n^2C \left( t_1^2 + t_2^2 - 2|K|t_1t_2 \right) \right].
\end{equation} 
For uncorrelated photon frequencies, i.e. for uncorrelated local environments we 
have $K=0$ and the trace distance decreases monotonically, corresponding to 
Markovian dynamics. However, as soon as the frequencies are anticorrelated, 
$K<0$, the trace distance is non-monotonic which signifies quantum memory 
effects and non-Markovian behavior. 

\subsection*{Consecutive interactions with local environments}

In our first experiment the local decoherence induced by the quartz 
plates in arm $1$ and arm $2$ act consecutively, and we change the size of 
the initial correlations between the local reservoirs by tuning the spectral width of 
the pump pulse. The results for the open system's trace distance dynamics are 
displayed in Fig.~2, which shows how the initial environmental correlations induce 
and influence the quantum non-Markovianity. First, when only the quartz plate in 
arm $1$ is active, the trace distance decreases monotonically demonstrating that
information flows continuously from the system to the environment. When
subsequently the quartz plate in arm $1$ becomes inactive and the quartz
plate in arm $2$ active, the trace distance increases again highlighting 
non-Markovian behavior and a reversed flow of information from the environment 
back to the open system. It is important to note that this increase of the trace 
distance implies a revival of the entanglement between the photon polarizations.
Whilst there exists earlier literature, e.g., on how the entanglement is transferred from the open system to the
environment~\cite{Lopez2008}, or on entanglement dynamics when the qubits interact with local non-Markovian
reservoirs~\cite{Bellomo}, our results are fundamentally different. We demonstrate that the entanglement can revive due to the nonlocal character of the dynamical map despite of the fact that the local interaction of the qubits with their environments is completely Markovian.
A further important aspect of the experimental scheme is that it enables to
measure the frequency correlation coefficient $K$ of the photon pairs by fitting the 
theoretical prediction of Eq.~(\ref{D}) to the experimental data. The fits shown
in Fig.~2 corresponding to panels (a)-(d) yield the values $K=-0.92$, $-0.66$, 
$-0.55$, and $-0.17$. Thus we see that the open system (polarization degrees
of freedom) functions as a quantum probe which allows us to gain nontrivial 
information on the correlations in the environment (frequency degrees of freedom).
To the best of our knowledge, our results present the first quantitative measurement of the correlation coefficient without reconstructing the joint frequency distribution.

\subsection*{Classical vs. quantum initial correlations between the local environments}

The open system dynamics of Eq.~(\ref{rho12}) depends solely on the
decoherence functions $\kappa_1$, $\kappa_2$, $\kappa_{12}$ and 
$\Lambda_{12}$ which in turn are completely determined by the Fourier transform
\eqref{Gt1t2} of the joint frequency distribution $P(\omega_1,\omega_2)$. 
Therefore, also the dynamical map of the open system depends only on $P(\omega_1,\omega_2)$ and not on the probability amplitudes $g(\omega_1,\omega_2)$.
This fact leads to the conclusion that the nonlocal memory effects,  and the revival of entanglement observed above,
do not depend on the specific phase relations between different frequency components, and consequently, whether the initial frequency correlations between the photons are of quantum or classical character.
To test experimentally whether the nature of the correlations plays a role in inducing the memory effects, we insert quartz plates also on the path of the pump 
pulse, thereby changing the dispersion relations of the pump pulse 
while keeping fixed the correlation coefficient $K$. The results in 
Fig.~3 show that the trace distance dynamics remains the same in this case and 
only depends on the correlation coefficient $K$ of the photon frequencies. Thus 
we see that nonlocal memory effects and the revival of the entanglement between 
the polarization states of the photons can be induced by purely classical 
correlations between local environments of the open system. 

\begin{figure}[tb]
\centering
\includegraphics[width=0.5\textwidth]{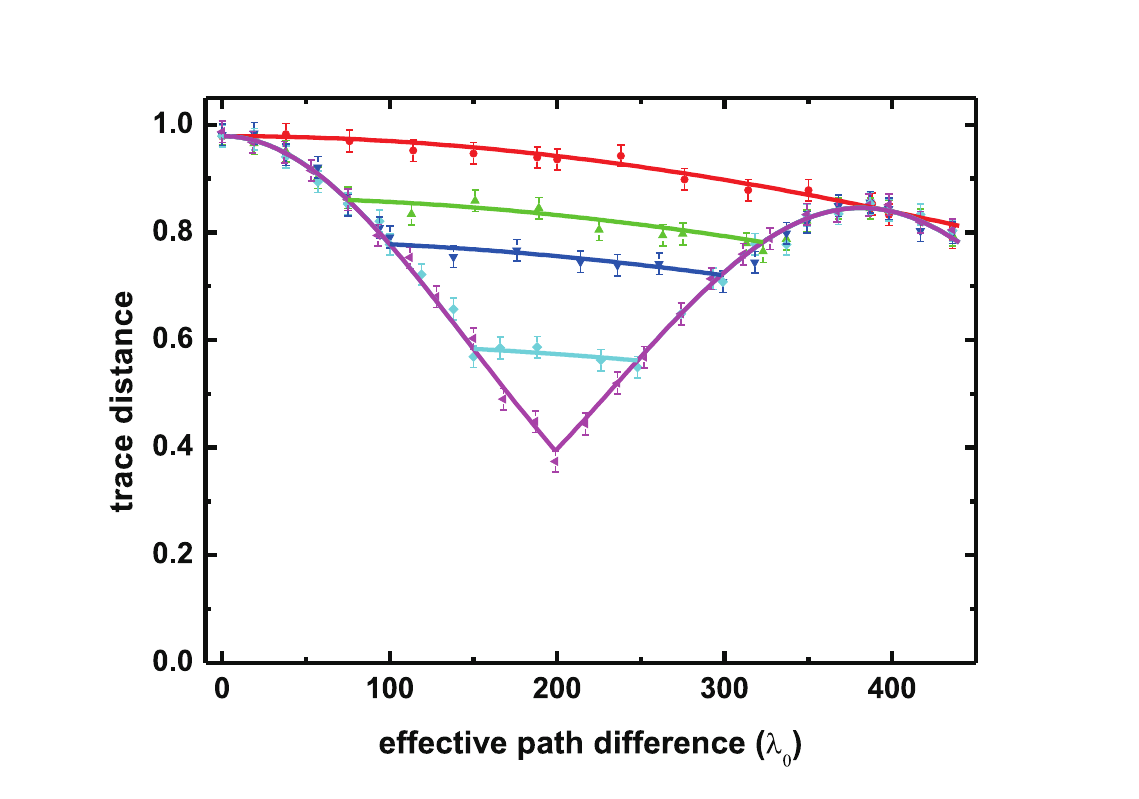}
\caption{\label{Fig:4} 
Trace distance dynamics for five different quartz plate configurations and a fixed
correlation coefficient of $K=-0.92$. For all curves the maximal length of the quartz 
plates in both arms is $199\lambda_0$, giving a maximal total effective path 
difference of $398\lambda_0$. From top to bottom, the curves show the results for 
the cases, when we start to add quartz plates to arm $2$ when the length of the 
plate in arm $1$ has reached the length of $0, 75, 100, 150, 199\lambda_0$ 
respectively, see the Methods Section for details. Hence, in the top curve the quartz plates are added simultaneously to 
both arms while in the bottom curve they act consecutively. The results 
demonstrate that the final state is the same while in terms of the information flow 
the behaviour is completely different. For the top (red) curve information flows 
slowly out of the system with a small rate and, in fact, the decoherence would 
completely halt in the ideal ideal case of fully anticorrelated frequencies 
($K=-1.0$). In the other extreme, for the bottom (violet) curve information first 
flows out of the system with a high rate and later on returns back nearly 
completely indicating strong memory effects. The solid lines are fits on the basis of 
Eq.~(\ref{D}), see the Methods Section for details}
\end{figure}

\subsection*{Different interaction configurations}

In the previous experiments the local dephasing environments acted 
consecutively. In our final experimental series we reposition the quartz plates in 
arm 2 so that eventually the dephasing is active simultaneously in both 
arms (see the Methods Section). The results are shown in Fig.~4.
We observe that the final state of the open system is always the same when the 
total effective path difference is equal to $398\lambda_0$. However, the amount of 
non-Markovianity and the flow of information from the environment to the open 
system decreases with increasing overlap of the action of the quartz plates. This 
demonstrates that initial correlations between the local environments and the 
reordering of the quartz plates allow to engineer the information flow between local 
systems while keeping fixed the final state. 
It is also important to note that when the local dephasing channels act 
simultaneously on our open bipartite system (the top red curve in Fig.~4), the initial 
correlations between the composite environments protect the entanglement 
between the qubits. One can see this also from Eq.~(\ref{D}) which
shows that for $t_1=t_2$ and $K=-1$ the trace distance remains constant and 
that there is thus no outflow of information from the system for the considered 
Bell-states. 

\section*{Discussion}

We have realized nonlocal quantum dynamical maps
of a photonic system and demonstrated experimentally how initial quantum or 
even classical correlations between the local environments of a composite open 
system induce nonlocal memory effects. While the global dynamics of the open 
system is non-Markovian, its local subsystems behave perfectly Markovian. 
The observation of the open system dynamics yields also important information 
about the environment. In fact, we have seen that a measurement
of the non-Markovian evolution of the two-photon polarization state leads to a 
novel method for the determination of the correlation coefficient of the photon 
frequencies. Thus, the open system dynamics can be viewed as a non-Markovian 
quantum probe which enables to extract nontrivial information on characteristic 
features of the environment.
This fact as well as our demonstration that initial correlations in the 
environment can protect the entanglement within an open system,
could find important applications in quantum technologies and control strategies, 
and in the development of novel diagnostics tools for complex quantum systems. 

\section*{Methods}

\subsection*{Controlling the widths of the pump pulse}
The width of the pump spectrum controls the frequency anticorrelations between 
the pair of photons created in the down conversion process.
By using filters and fused silica plates, we can control the width of the pump 
pulses which are depicted in Fig.~5. The laser source is filtered to 
$3\,\rm{nm}$ (FWHF) and then passes through the frequency doubler
($1.5\,\rm{mm}$ thick BiB$_3$O$_4$). Hence, the bandwidth of the UV pulse is 
about $0.52\,\rm{nm}$ as shown in Fig.~5 (b). To obtain a sharper
spectrum, as shown in Fig.~5 (a),  we insert a thin fused silica plate, which is 
$0.05\,\rm{mm}$ thick and coated with a partial reflecting coating on each side, 
with approximately $75$\% at $390\,\rm{nm}$. For the spectra displayed in Fig.~5 
(c) and (d), we have no filter before the doubler and no fused silica plate after the 
doubler, and for Fig.~5 (d) we also replace the doubler by a 
$0.3\,\rm{mm}$-thick BBO.

\subsection*{Counting statistics}
The details of the counting statistics and pump power for the experimental results 
of Fig.~2 are as follows. For Fig.~2 (a) the pump power is about 
$2.4\,\rm{mW}$ and the total coincident count rate 
on the basis vectors
HH, HV, VH, VV is about $18,000$ in $10$ seconds, (b) $10\,\rm{mW}$ and the 
rate is about $35,000$ in $4$ seconds, (c) $20\,\rm{mW}$ and the rate is 
about $35,000$ in $2$ seconds, and (d) $20\,\rm{mW}$ and rate is about 
$36,000$ in $4$ seconds. Please note that the coincident count rate always 
depends also on the bandwidth of UV pump source.

\subsection*{Fitting the experimental data}
Employing Eq.~(3), the theoretical fits of the 
experimental data presented in Figs.~2 and 3 have been
carried out as follows. We first use the function 
$f(x) = A\exp\left(-Bx^2\right)$ to make a fit for the process when the quartz plates 
are added to arm 1. From this process, we can find 
$A$ and $B$ which agree with the experimental data. Then in the second part, 
when the quartz plate in arm 2 is active, we use the function
$g(x)=A\exp\left[-B\left(199^2+(x-199^2)-2|K|199 (x-199)\right)\right]$. Here, $199$ 
corresponds to the effective path difference after the first quartz plate. Since $A$ 
and $B$ are determined by the first fit, we can easily determine $K$ from the 
second fit.

For Fig.~4, again keeping in mind Eq.~(3), we can do the fitting in the following way.
First, we use the functions $f(x)$ and $g(x)$ mentioned above to make a fit of the 
lowest curve and determine $A_1$, $B_1$, and $K_1$. Then we use a function 
$G(x) = A\exp\left[ -B \left( x^2/2 - |K| x^2/2\right) \right]$ to fit the top curve of 
the figure. From here, we determine $A_2$, $B_2$, and $K_2$. Then we 
determine $A_0=(A_1+A_2)/2$, $B_0=(B_1+B_2)/2$, and $K_0=(K_1+K_2)/2$ 
which are then used to plot the functions
$F(t_1,t_2)= A_0\exp\left[ -B_0 \left(t_1^2 + t_2^2-2|K_0|t_1t_2\right)\right]$ in the 
figure.

\subsection*{Trace distance dynamics with different dispersions of the pump pulse}
For the experimental results of Fig.~3,  
the FWHM of the pump pulse is about $0.73\,\rm{nm}$, the pump power is about 
$20\,\rm{mW}$, and the total coincident count rate on the basis 
vectors HH, HV, VH, VV is about $35,000$ in $2$ seconds.

\begin{figure}[t]
\centering
\includegraphics[width=0.5\textwidth]{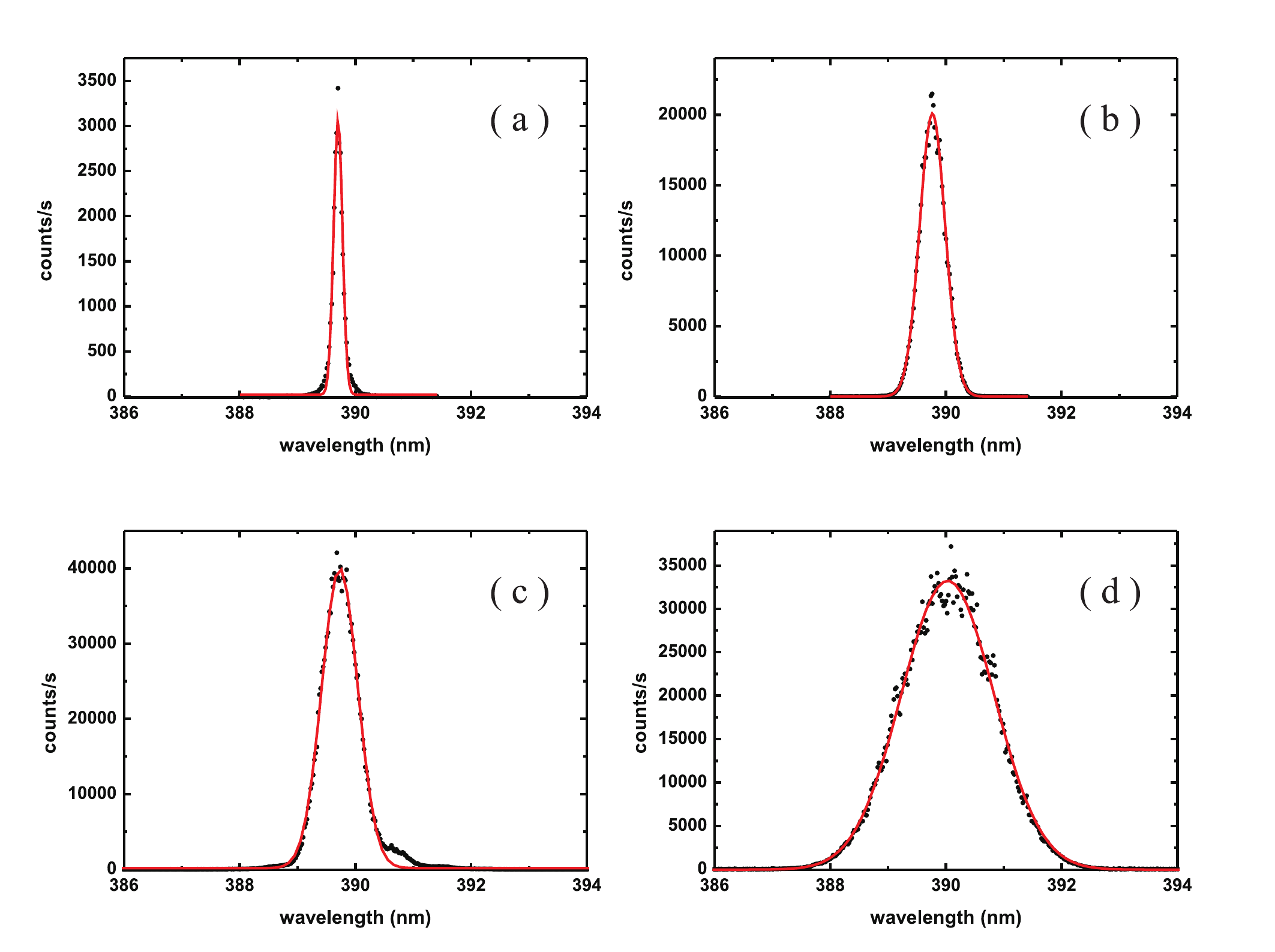}
\caption{\label{Fig:S1} 
The frequency spectra of the ultra-violet (UV) pump pulses used
for the down conversion process. Full widths at half maximum (FWHM) of the 
pulse spectra are 
(a) $\delta=0.18\,\rm{nm}$,
(c) $\delta=0.52\,\rm{nm}$,
(c) $\delta=0.73\,\rm{nm}$, and
(d) $\delta=1.89\,\rm{nm}$.
The solid lines represent Gaussian fits of the experimental data.}
\end{figure}

\subsection*{Quartz plate configurations}
Figure 6 shows the five different quartz plate configurations used for Fig.~4. 
To obtain the data of Fig.~4, the FWHM of the UV pulse is about $0.18\,\rm{nm}$, the pump power is 
about $2.4\,\rm{mW}$, and the total coincident count rate on the basis 
vectors HH, HV, VH, VV is about $18,000$ in $10$ seconds. 

\begin{figure}[t]
\centering
\includegraphics[scale=0.45]{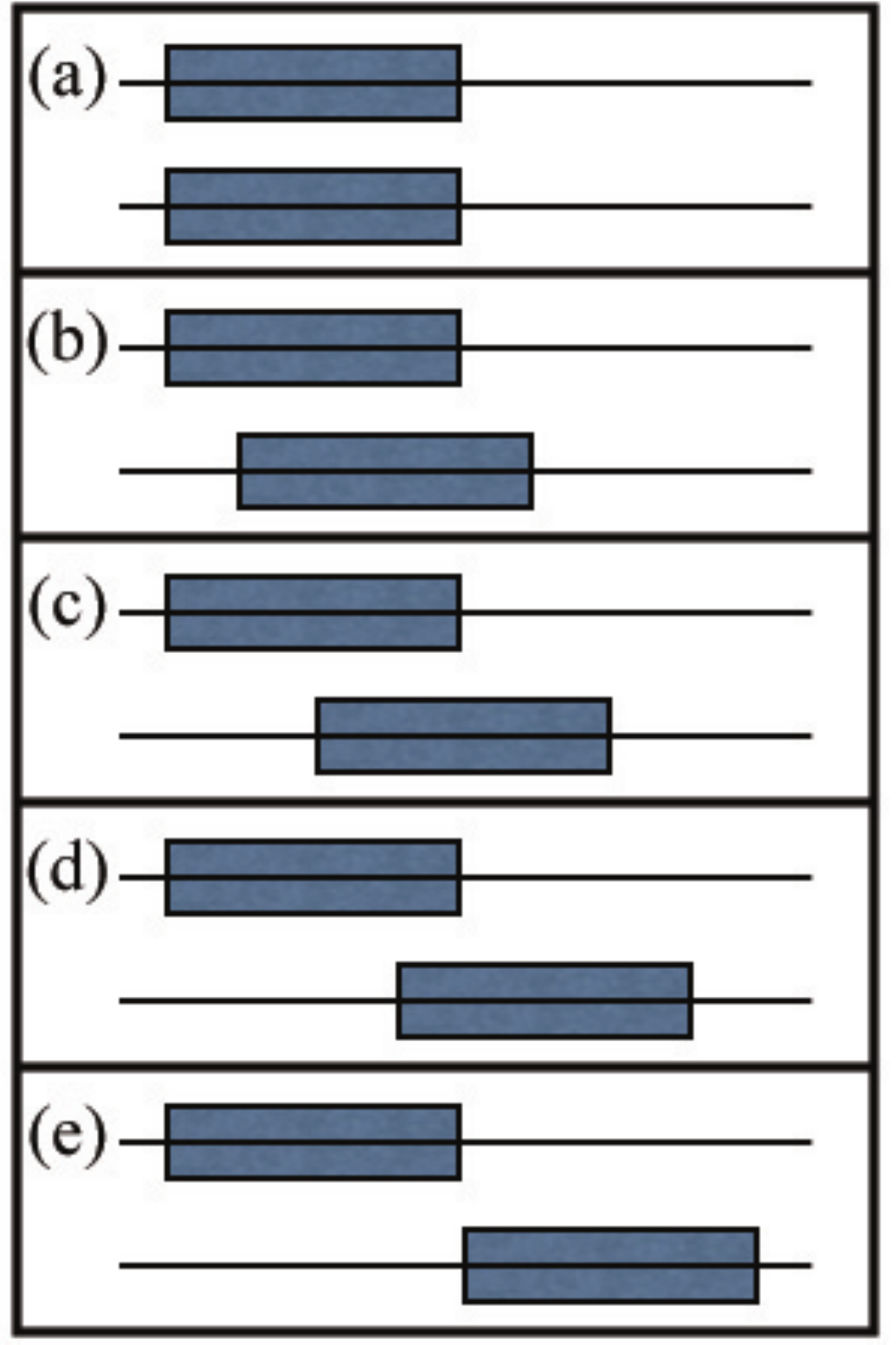}
\caption{\label{Fig:S2} 
The schematic view of the different quartz plate configurations.
The configurations from (a) to (e) correspond to the curves of Fig.~4  from top to bottom. For all cases the maximal length of the quartz 
plates in both arms is $199\lambda_0$, giving a maximal total effective path 
difference of $398\lambda_0$. We start to add quartz plates to arm $2$ when the 
length of the plate in arm $1$ has reached the length of
(a) $0$,
(b) $75\lambda_0$,
(c) $100\lambda_0$,
(d) $150\lambda_0$, and
(e) $199\lambda_0$.
}
\end{figure}

\section*{Acknowledgments}
This work was supported by the National Natural Science Foundation of China (Grant Nos.~10874162, 11074242, 11104261 and 60921091), the National Basic Research Program of China, the Foundation for the Author of National Excellent Doctoral Dissertation of PR China (Grant No.~200729), the Fundamental Research Funds for the Central Universities (Grant 
Nos.~2030020004 and 2030020007), the Anhui Provincial Natural Science Foundation (Grant No.~11040606Q47), the Academy of Finland (mobility from Finland 259827), the Magnus Ehrnrooth Foundation (Finland), the Jenny and Antti 
Wihuri Foundation (Finland), the Graduate School of Modern Optics and 
Photonics (Finland), and the German Academic Exchange Service (DAAD).

\section*{Author Contributions}
B.-H.L., D.-Y.C, Y.-F.H., C.-F.L., and G.-C.G. planned, designed and implemented 
the experiments.
C.-F.L, G.-C. G., E.-M.L., H.-P.B., and J.P. carried out the theoretical analysis and 
developed the interpretation.
B.-H.L., C.-F.L., E.-M.L, H.-P.B., and J.P. wrote the paper and all authors 
discussed the contents.

\section*{ Author Information}
The authors declare that they have no competing financial interests.
Correspondence and 
requests for materials should be addressed to C.-F.L. or J.P.

\end{document}